\documentclass[aps,prb,preprint]{revtex4}
\usepackage{latexsym}
\usepackage{graphicx}
\usepackage[arrowdel,italicdiff]{physics}
\usepackage{multirow}
\usepackage{Tabbing}
\usepackage{subfigure}    
\usepackage{amsmath,amsfonts,amsthm,amssymb}
\usepackage{epsfig}

\usepackage{graphicx,float,wrapfig}
\usepackage{float}
\usepackage{hyperref}
\usepackage{setspace}
\usepackage{fancyhdr}

\usepackage{extramarks}
\usepackage{chngpage}
\usepackage{soul,color}
\usepackage{afterpage}

\usepackage{lastpage}

\usepackage{placeins}

\newcommand{\figwidth}{3. in}

\newcommand{\figwtwice}{4.3 in}

\newcommand{\trm}[1]{\textrm{#1}}
\newcommand{\ti}[1]{\tilde{#1}}

\newcommand{\iim}{\textrm{i}}

\usepackage{siunitx}
\usepackage{mathdots}
\usepackage{mathrsfs}
\usepackage{tabto}
\usepackage{enumerate}
\usepackage{url}
\usepackage{booktabs}

\begin{document}

\title{
  Imaginary-time time-dependent density functional theory and its application for robust convergence of electronic states
}
\author{Cedric Flamant$^{(1)}$}
\author{Grigory Kolesov$^{(2)}$}
\author{Efstratios Manousakis$^{(3,4)}$}
\author{Efthimios Kaxiras$^{(1,2)}$}
\affiliation{
$^{(1)}$Department of Physics, Harvard University, Cambridge, Massachusetts 02138, USA\\
$^{(2)}$John A. Paulson School of Engineering and Applied Sciences, Harvard University, Cambridge, Massachusetts 02138, USA\\
$^{(3)}$ Department  of  Physics and National High Magnetic Field Laboratory,
  Florida  State  University,  Tallahassee,  Florida  32306-4350,  USA\\
$^{(4)}$Department   of    Physics,   University    of   Athens,
  Panepistimioupolis, Zografos, 157 84 Athens, Greece
}
\date{\today}

\begin{abstract}
  Reliable and robust convergence to the electronic ground state within density functional theory (DFT) Kohn-Sham (KS) calculations remains a thorny issue in many systems of interest. In such cases, charge sloshing can delay or completely hinder the convergence. Here, we use an approach based on transforming the time-dependent DFT
  equations to imaginary time, followed by imaginary-time evolution, as a reliable alternative to the self-consistent field (SCF) procedure
  for determining
  the KS ground state. We discuss the theoretical and technical
  aspects of this approach and show that the KS ground state
  should be expected to be the long-imaginary-time output of the evolution, independent of the exchange-correlation functional or the level of theory used to simulate the system.  By maintaining self-consistency between the single-particle wavefunctions and the electronic density throughout the determination of the stationary state, our method avoids the typical difficulties encountered in SCF.
  To demonstrate dependability of our approach, we apply it to selected systems which struggle to converge with SCF schemes. In addition, through the van Leeuwen theorem, we affirm the physical meaningfulness of imaginary time TDDFT, justifying its use in certain topics of statistical mechanics such as in computing imaginary time path integrals.
  \end{abstract}
  \maketitle

  \section{Introduction} \label{sec:intro}

  Density functional  theory (DFT)  is a  widely used approach
  enabling \emph{ab initio} calculations  of
electronic  and  material  properties.  Unlike  direct  approaches  to
studying quantum systems through  the Schr\"odinger equation where the
wavefunction is the central object, DFT uses
the electron density $n(\vb{r})$ as the fundamental
physical quantity.  In   principle,  through  the  Hohenberg-Kohn
theorem\cite{hohenbergkohn},  the  ground state $n(\vb{r})$ of a  system  uniquely
determines all of its observables. It  is standard practice to use the
Kohn-Sham (KS) system\cite{kohnsham} of non-interacting fermions as a shortcut to obtaining
the ground  state density,  employing specially  formulated potentials\cite{ceperleyCA,PBE}
that   are   functionals   of  $n(\vb{r})$ to   approximate   the
electron-electron interactions.

There are many techniques to find the ground state of the KS
equations, including methods which: (a) aim at direct determination
of the minimum of the KS total energy functional\cite{dirmin1,dirmin2,dirmin3}; and (b) use iterative
methods based on diagonalization of the
KS Hamiltonian in conjunction with iterative improvements of the
ground state charge density through mixing. The present work is 
distinct from both these approaches. We focus on comparing our method to type (b) approaches.

For a system with $N$ electrons, the lowest $N$ eigenstates to the KS equations determine
$n(\vb{r})$, which itself appears in the KS equations through an effective single-particle potential. In general,
finding  the  set  of  $N$  eigenstates   that  satisfy the KS
equations    involves  an
iterative process known as self-consistent field  (SCF) iterations
that produce successively better approximations to the solution.
In its simplest conceptualization the iterative  approach involves
solving the  eigenvalue problem
for an initial density distribution, then using the resulting  eigenstates to
produce  the  next  approximation to the density.
When this approach is iterated, except for the simplest systems, it  rarely   converges  to  a
self-consistent solution. In order to  stabilize the SCF loops and improve
the convergence rate,  various  mixing schemes  are typically employed.  These   schemes  take  advantage  of   the  information
contained in  multiple previous trial  densities to select the
next one. A popular mixing scheme is direct inversion of the iterative subspace (DIIS), also known as  Pulay
mixing\cite{PULAY1,PULAY2}.

When SCF schemes  require many  iterations  to reach  an
acceptable solution, or fail to converge, the choices are to
change the mixing scheme  or its parameters, start with a
different  density, or fractionally occupy states\cite{Michelini,Rabuck} which some methods implement by introducing a fictitious  electronic temperature  (Fermi
smearing\cite{mermin_fermi,fermismearing}). If these fail, one can resort to computationally-intensive direct
minimization methods\cite{dirmin1,dirmin2,dirmin3} to find a solution. The convergence difficulties for SCF usually arise in  systems with large unit cells and
in metallic  systems\cite{Anglade}, or when an excited state is desired. The  small differences in  eigenenergies of
the single-particle  states, as  well as the  presence of  many states
near the  Fermi level, can  cause very different eigenstates  to be occupied
from step  to step.  This can lead  to large  variations in  the density,
causing the phenomenon  known as charge sloshing\cite{Kresse-charge-sloshing}
where a fluctuating charge density from step to step is observed with insufficient attenuation to reach convergence.

In the present paper we transform the time-dependent KS (TDKS) equations of time-dependent density functional
theory (TDDFT)\cite{TDDFT,runge_gross,vanleeuwen}  to imaginary time\cite{KKM}. We use these equations to propagate an initial
state to very long imaginary time, refining it down to the KS state corresponding to its lowest energy component. The idea of using imaginary-time propagation (ITP) to find eigenstates is well-known, and it is frequently used to find ground state solutions to the Schr\"odinger equation describing single-particle systems with a fixed potential\cite{Bader,Lehtovaara}. Imaginary time steps have also been used to find self-consistent solutions to the Hartree-Fock equations\cite{Davies} and for nuclear energy density functional calculations\cite{Ryssens}. It has also been employed in a DFT context as an alternative to the diagonalization step to find the single-particle KS eigenstates for a fixed electronic density\cite{itp1,itp2}. However, imaginary-time evolution has yet to be examined as a stand-alone substitute to iterative density updating in solving the KS equations. In the present method both the density and wavefunction evolve together towards the ground state according to the imaginary time TDKS equations, remaining consistent with each other throughout the calculation. We discuss the theoretical foundation of the imaginary-time evolution of the
KS system, a procedure which is non-unitary, 
requiring re-orthonormalization of the states at each imaginary time-step.
We show that the proof provided by van
Leeuwen\cite{vanleeuwen} for TDDFT can be extended to imaginary-time TDDFT (it-TDDFT), affirming in principle that the density of a KS system will evolve in imaginary time in the same manner as the true many-body interacting system. The imaginary-time propagation method in DFT has attractive theoretical and practical benefits when applied to systems that are challenging to study using standard  methods of solving the KS equations, as we demonstrate on model systems.

We benchmark our approach by applying it to the benzene molecule
and show that it converges to the same ground state energy as other
SCF-based methods. Next, we apply our method
to systems with known difficulties in achieving
convergence. We chose to examine a copper nanocluster Cu$_{13}$
with fixed magnetization and a spin-unpolarized Ru$_{55}$ nanocluster. We show that self-consistent solutions are hard to realize in both
systems using the most popular standard approach, SCF with Pulay mixing.
In general, we find that while requiring more computation, our method is more 
dependable and more autonomous compared to SCF. It provides a good alternative
to existing methodologies when the latter fail to converge in challenging systems,
or if a user wishes to find an unfamiliar system's ground state with minimal intervention; this can be particularly useful when computations are carried out in an automated fashion on large clusters of processors. 

The paper is organized as follows: Section~\ref{section:method} presents the method, Section~\ref{section:considerations} extends the van Leeuwen theorem to imaginary time and discusses certain theoretical considerations that establish the method's robustness, Section~\ref{section:calculations} gives some example calculations, and Section~\ref{section:conclusion} contains our conclusions.


\section{Methodology}
\label{section:method}

\subsection{Imaginary-Time Propagation} \label{sec:ITP}

First,  let  us take  the  Hamiltonian $\hat{H}$ to be  time-independent.
Under the  substitution $t \to - \iim  \tau$, where $\tau$ is real, the time evolution  operator transforms from $e^{-\iim t\hat{H}}$ to
$e^{-\tau  \hat{H}}$.   When  $\ket{\Phi_i}$   is  an   eigenstate  of
$\hat{H}$, the time evolution of the eigenstate switches from rotating
its phase proportionally to its energy:
\begin{align}
  \ket{\Phi_i(t)}     =     e^{- \iim t\hat{H}}\ket{\Phi_i}    =     e^{- \iim t
    E_i}\ket{\Phi_i},
\end{align}
 to  shrinking its  amplitude by  an  exponential factor  with a  rate
 proportional to the energy:
\begin{align}
  \ket{\Phi_i(\tau)}   =   e^{-\tau\hat{H}}\ket{\Phi_i}   =   e^{-\tau
    E_i}\ket{\Phi_i}.
\end{align}
For the  case of a  time-dependent Hamiltonian the  previous equations
still hold  for infinitesimal  amounts of time  $\Delta t$  or $\Delta
\tau$.   For   an  arbitrary  initial   wavefunction  $\ket{\Psi(0)}$,
imaginary-time propagation amounts to
\begin{align}
  \ket*{\ti{\Psi}(\tau)}   =   \sum_{i=0}^{\infty}  A_i(0)   e^{-\tau   E_i}
  \ket{\Phi_i},
  \label{eq:interactingevolution}
\end{align}
where $A_i(0)$ is the amplitude  of the eigenstate component initially
present. As imaginary time goes to infinity, $\tau \to \infty$, the eigenstate $\ket{\Phi_j}$ corresponding to  the  lowest  energy eigenvalue with $A_j(0) \neq 0$ will  dominate.  We  can choose  to keep  the state
$\ket{\Psi(\tau)}$ normalized  by dividing  by the  norm $\Omega(\tau)
\equiv \sqrt{\braket*{\ti{\Psi}(\tau)}{\ti{\Psi}(\tau)}} = \sqrt{\sum_{i=0}^{\infty}     \abs{A_i(0)}^2 e^{-2\tau E_i}}$,

\begin{align}
  \ket{\Psi(\tau)}  = \sum_{i=0}^{\infty}\frac{  A_i(0) e^{-\tau  E_i}
  }{\Omega(\tau)}\ket{\Phi_i},
  \label{eq:interactingevolutionnormalized}
\end{align}
which   then   yields   $\lim_{\tau\to  \infty}   \ket{\Psi(\tau)}   =
\ket{\Phi_j}$. Note that the state $\ket{\Psi(\tau)}$ could refer to a single-particle
wavefunction $\Psi(\vb{r})$ or a many-body wavefunction $\Psi(\vb{r}_1,\ldots,  {\vb{r}}_N)$; the above discussion is applicable to either case. Since an arbitrary initial  state generated by randomizing  the coefficients in
some basis  is likely to  have a nonzero ground  state component,
ITP is often used to find ground state wavefunctions and energies.

\subsection{Implementation within the Kohn-Sham Formalism}

In TDDFT, starting from an initial
state, the KS system obeys the equations of
motion (in atomic units):
\begin{subequations}
\begin{align}
  \iim \pdv{}{t}  \phi_j( {\vb{r}},t) &= \hat{H}_{\trm{KS}}[n({\vb{r}}, t)]
  \phi_j\qty({\vb{r}},t), \\  \hat{H}_{\trm{KS}}[n({\vb{r}},  t)] &\equiv
  \qty[ -\frac{ \laplacian}{2 } + v_s({\vb{r}},t)],
\end{align}
  \label{eq:TDKS}
\end{subequations}
with time-dependent effective potential
\begin{align}
  v_s[n({\vb{r}},t)]  =   v({\vb{r}})  +   v_{\trm{H}}\qty[n({\vb{r}},t)]  +
  v_{\trm{xc}}\qty[n\qty({\vb{r}},t)].
  \label{eq:TDKS-effective}
\end{align}
In these expressions, $v({\vb{r}})$ is the external potential and
\begin{align}
  v_{\trm{H}}\qty[n({\vb{r}},t)]       &=       \int       \dd{{\vb{r}}'}
  \frac{n\qty({\vb{r}}',t)}{\abs{{\vb{r}}                                -
      {\vb{r}}'}}, \label{eq:TDKS-Hartree}\\         v_{\trm{xc}}\qty[n\qty({\vb{r}},t)]        &=
      \fdv{E_{\trm{xc}}\qty[n({\vb{r}}, t)]}{n\qty({\vb{r}},t)}, \label{eq:TDKS-exc}\\
  n({\vb{r}},t)   &= \sum_{j=1}^{N} \abs{\phi_j\qty({\vb{r}},t)}^2.
  \label{eq:TDKS-density}
\end{align}

The  Kohn-Sham  time-evolution  can  be reformulated  in  terms  of  a
time-propagator which acts on single-particle states and is given by
\begin{align}
  \ket{\phi_j(t)} &=  \hat{U}\qty(t,t_0) \ket{\phi_j(t_0)},  \\ \quad
  \hat{U}\qty(t,t_0) &= \hat{\mathcal{T}}\exp\qty(- \iim  \int_{t_0}^{t}
  \hat{H}_{\trm{KS}}[n({\vb{r}},t')] \dd{t'}),
\end{align}
where $\hat{\mathcal{T}}$ is the time-ordering operator.  In imaginary
time, applying the substitution $t \to -\iim \tau$ results in
\begin{align}
  \ket{\phi_j(\tau)}         &=         \hat{\mathcal{U}}\qty(\tau,\tau_0)
  \ket{\phi_j(\tau_0)},  \\  \quad  \hat{\mathcal{U}}\qty(\tau,\tau_0)  &=
  \hat{\mathcal{T}}_\tau        \exp\qty(-        \int_{\tau_0}^{\tau}
  \hat{H}_{\trm{KS}}[n(\vb{r},\tau')] \dd{\tau'}),
\end{align}
where $\hat{\mathcal{T}}_\tau$ now time-orders in imaginary time. Note
that the imaginary-time propagator is not unitary.  

Employing the same numerical scheme  used for real time propagation of KS
states on an atomic basis\cite{kolesov2015real}, we evolve in imaginary-time the single-particle states using
finite time steps $\Delta \tau$ and we approximate the instantaneous imaginary-time
propagator with the second-order Magnus expansion:
\begin{align}
  \hat{\mathcal{U}}\qty(\tau   +   \Delta   \tau,  \tau)   &\approx   \exp
  \qty[-\Delta   \tau   \hat{H}_{\trm{KS}}\qty(\tau   +   \frac{\Delta
      \tau}{2})],\\  \hat{H}_{\trm{KS}}  (\tau   )  &\equiv
      \hat{H}_{\trm{KS}}[n({\vb{r}}, \tau )].
\end{align}

The Hamiltonian at the midpoint is  approximated as the average of the
Hamiltonians        at        $\tau_i$        and        $\tau_{i+1}$,
$\hat{H}_{\trm{KS}}\qty(\tau_i   +   \frac{\Delta  \tau}{2})   \approx
\frac{1}{2}\qty[\hat{H}_{\trm{KS}}(\tau_i)                           +
  \hat{H}_{\trm{KS}}(\tau_{i+1})]$.   Each   step   is   iterated   to
self-consistency  in  order   to  make  use  of   the  Hamiltonian  at
$\tau_{i+1}$. We  use the Pad\'e rational  polynomial approximation of
arbitrary  degree to  obtain the  general matrix  exponential. Further
details  of the  numerical propagation  can be  found in
our earlier work\cite{kolesov2015real}, which describes TDAP-2.0, a TDDFT code we used,
built on top of SIESTA\cite{siesta02}, a DFT package which uses strictly localized
basis sets. While the  midpoint Hamiltonian greatly aids stability
and energy conservation in real  time propagation, in practice we have
found that  for imaginary-time propagation  we can just use  the first
step   in  the   iterative   procedure,  which   simply  applies   the
approximation $\hat{H}_{\trm{KS}}\qty(\tau_i  + \frac{\Delta \tau}{2})
\approx  \hat{H}_{\trm{KS}}\qty(\tau_i)$.  This explicit propagation
is faster  since  the
Hamiltonian only needs to be  evaluated once per propagation step, and
the  effect  on  the  size  of the  maximum  stable  time-step  appears
negligible compared to the implicit method using the midpoint Hamiltonian. 
This is expected since imaginary-time propagation is
inherently more stable than the real-time propagation the TDAP-2.0
code was originally designed to solve.

Because   the  imaginary-time   propagator  is   not  unitary,   the
single-particle states lose their normalization and generally cease to
be  orthogonal.  The  simple  expression  for density  in
Eq.~(\ref{eq:TDKS-density}) becomes  more complicated  if the
single-particle  states  $\phi_j$  are   non-orthonormal. It is
convenient to  reorthonormalize the  single-particle states at each time
step. The details of  how the orthogonalization
is achieved  do not  affect the  physics, as we show  in Section~\ref{ssec:orthon}.  We  use  the modified  Gram-Schmidt  algorithm  to
orthonormalize the states.

While we employ a localized atomic basis for our calculations, the method we propose
is independent of the basis used to represent the Kohn-Sham orbitals, and can easily be implemented in other popular bases, like plane waves or gaussians.

\section{Theoretical Considerations}
\label{section:considerations}

\subsection{Van Leeuwen Theorem in Imaginary Time} \label{ssec:vlt}

The van Leeuwen theorem states that a  time-dependent particle density
$n({\vb{r}},t)$  belonging to  a many-particle  system with  two-particle
interaction $\hat{W}$ can  always be reproduced by a unique  (up to an
additive   purely   time-dependent    constant)   external   potential
$v'({\vb{r}},t)$ in  another many-particle  system that uses  a different
two-particle interaction  $\hat{W}'$, under the mild  restriction that
the density has to be analytic  in time\cite{vanleeuwen}. If we choose the two-particle
interaction  in  this  other  system  to be $\hat{W}'  = 0$, the  theorem guarantees the  existence of
the  effective potential  $v_s({\vb{r}},t)$ for  a Kohn-Sham  system that
reproduces the  same time-dependent density as  the interacting system
of interest. Here we point out the modifications to the original theorem
in order to make it compatible with imaginary-time evolution.

A complex $t$ value does not pose any problems
with the operations performed in the original proof, where $t$ appears in some
time derivatives  but otherwise is treated as a  parameter. We add time-dependent uniform potentials $\lambda(t)$ and $\lambda'(t)$ to the unprimed and primed Hamiltonians to conserve the norm of the wavefunctions. The origin of these terms will be discussed in the next section.
The  Hamiltonian $\hat{H}$ of a finite many-particle system is then given by
\begin{align}
  \hat{H}(t) = \hat{T} + \hat{V}(t) + \hat{W} + \lambda(t),
\end{align}
expressed in terms of creation and annihilation operators
\begin{subequations}
\begin{align}
  \hat{T} &= - \frac{1}{2} \int \dd[3]{\vb{r}} \hat{\psi}^\dag(\vb{r})
  \laplacian \hat{\psi}(\vb{r}), \\  \hat{V}(t) &= \int \dd[3]{\vb{r}}
  v(\vb{r}t)                               \hat{\psi}^\dag\qty(\vb{r})
  \hat{\psi}(\vb{r}), \label{eq:Vdef}\\ \hat{W} &= \int \dd[3]{\vb{r}}
  \dd[3]{\vb{r}'}       w\qty(\abs{      \vb{r}       -      \vb{r}'})
  \hat{\psi}^\dag(\vb{r}) \hat{\psi}^\dag(\vb{r}') \hat{\psi}(\vb{r}')
  \hat{\psi}({\vb{r}}).
\end{align}
\end{subequations}
Since $\lambda(t)$ is not an operator,  it does not affect any of  the commutators involving
$\hat{H}(t)$  in  the various Heisenberg   equations  of  motion underpinning the proof of the van Leeuwen theorem. There is only one detail to
note,  regarding  the  freedom  to  add an  arbitrary  $C(t)$  to  the
potential  of  the  primed  system,  $v'({\vb{r}},t)$,  in  the  original
proof. From Eq.~(\ref{eq:Vdef}) a time-dependent
constant in the potential modifies the Hamiltonian by an additional term $C(t)\hat{N}$,
where $\hat{N}$ is  the number operator. For the  systems of interest,
the  number of  particles $N$ is fixed so $C(t)N$
is a time-dependent uniform potential like $\lambda'(t)$, which means  that a
norm-conserving  $\lambda'(t)$ will  cancel any  effect from  the
choice of  $C(t)$. Thus, with $\lambda(t)$  and $\lambda'(t)$ 
chosen to  ensure that  the norm  of states in  both the  unprimed and
primed systems is held at unity, the  van  Leeuwen  theorem  holds  in
imaginary time. This  is a  powerful result since it allows us to think about imaginary-time propagation in the Kohn-Sham  system in  terms of  what it does  in the  real system,
allowing the Wick-rotation  connections from quantum mechanics
to statistical mechanics to be employed. For example, it justifies the
use of the  Kohn-Sham system as a stand-in for  the interacting system
in our calculations performed  for  imaginary time  path integrals\cite{KKM}.

\subsection{Maintaining Orthonormalization} \label{ssec:orthon}

Orthonormalization  of the  single-particle  states  is equivalent  to adding a purely time-dependent  function $\lambda(t)$ to the many-body
Hamiltonian. This takes  care of
holding  the  wavefunction normalized,  both  in  the interacting  and
Kohn-Sham  systems, as  well as  accounting for  the orthogonalization
step we use in the Kohn-Sham state propagation. 

We first consider the interacting system.  In real
time  propagation, the  choice  of $\lambda(t)$  does  not affect  the
dynamics  of density  since this  spatially-constant offset  in energy
only results in changing the phase of the wavefunction:

\begin{subequations}
\begin{align}
  \ket{\Psi(t)}  &=   \hat{U}\qty(t,t_0)
  \ket{\Psi(t_0)},\nonumber\\     \quad     \hat{U}\qty(t,t_0)     &=
   \hat{\mathcal{T}}\exp\qty(- \iim     \int_{t_0}^{t}    \hat{H}(t')     +
  \lambda(t')  \dd{t'})\nonumber  \\  &=  \hat U_{\lambda}(t,t_0)
  \hskip 0.05 in \hat{\mathcal{T}}  \hskip 0.05 in \hat U_{\hat H}(t,t_0),\\
  \hat U_{\lambda}(t,t_0) &\equiv
  \exp\qty(-  \iim \int_{t_0}^{t}
  \lambda(t') \dd{t'}), \\
  \hat U_{\hat H}(t,t_0) &\equiv
  \exp\qty(- \iim \int_{t_0}^{t}
  \hat{H}(t') \dd{t'}).
\end{align}
\end{subequations}
In  imaginary-time propagation, $\lambda(\tau)$ modifies
the  imaginary-time  propagator  $\hat{\mathcal U}\qty(\tau,\tau_0)$ by  a
time dependent magnitude,
\begin{subequations}
\begin{align}
  \hat {\mathcal U}\qty(\tau,\tau_0)  &=   \hat {\mathcal U}_{\lambda}(\tau,\tau_0)
   \hskip 0.05 in\hat{\mathcal{T}}_{\tau} \hskip 0.05 in \hat {\mathcal U}_{\hat H}(\tau,\tau_0), \label{eq:ITPwithC}\\
    \hat {\mathcal U}_{\lambda}(\tau,\tau_0) &\equiv
  \exp\qty(-  \int_{\tau_0}^{\tau}
  \lambda(\tau') \dd{\tau'}), \\
  \hat {\mathcal U}_{\hat H}(\tau,\tau_0) &\equiv
  \exp\qty(- \int_{\tau_0}^{\tau}
  \hat{H}(\tau') \dd{\tau'}).
\end{align}
\end{subequations}
If $\lambda(\tau)$  is arbitrary, the  norm of  the
wavefunction will change in  time, incorrectly scaling the expectation
values of observables like density  and energy. The norm of the wavefunction
can be held  fixed   by  choosing
$\lambda(\tau)$   to   counteract    the   norm-altering   effect   of
$\hat{\mathcal{T}}_{\tau}\exp\qty(-               \int_{\tau_0}^{\tau}
\hat{H}(\tau') \dd{\tau'})$ when it acts on $\ket{\Psi(\tau_0)}$. Note
that  such  a  $\lambda(\tau)$  will   also  depend  on  the  starting
state. For example, in the time-independent Hamiltonian case presented
in         Section~\ref{sec:ITP}, from Eq.~(\ref{eq:interactingevolutionnormalized})
\begin{align}
  \hat {\mathcal U}_{\lambda}(\tau,\tau_0)  =
  \qty[\sum_{j=0}^{\infty}   \abs{A_j(0)}^2   e^{-2\tau   E_j}]^{-1/2}
\end{align}
which implies
\begin{align}
 \lambda(\tau) =  \frac{1}{2} \dv{}{\tau}  \ln
  \qty[\sum_{j=0}^{\infty}   \abs{A_j(0)}^2   e^{-2\tau  E_j}]   =   -
  \ev{E(\tau)},
\end{align}
that is, to keep the wavefunction normalized, $\lambda(\tau)$ is such that
the energies of the Hamiltonian are measured relative to $\ev{E(\tau)}$.
This result holds more generally for time-dependent
Hamiltonians    as   well,    which    can   be    shown   by    using
$\hat{\mathcal{U}}\qty(\tau,\tau_0)$  from  Eq.~(\ref{eq:ITPwithC})  and
differentiating     the     norm-conserving      equation     $1     =
\bra{\Psi(\tau_0)}\hat{\mathcal{U}}^{\dag}\qty(\tau,                \tau_0)
\hat{\mathcal{U}}\qty(\tau,\tau_0)   \ket{\Psi(\tau_0)}$   to   solve   for
$\lambda(\tau)$. We will assume that such a $\lambda(\tau)$ is used in
the   interacting   system  so   that   the   system  always   remains
normalized. 

In the Kohn-Sham system the propagator is given by
\begin{align}
  \hat{\mathcal U}\qty(\tau,\tau_0)  &= \hat {\mathcal U}_{\lambda_{\trm{KS}}}(\tau,\tau_0)
  \hskip 0.05 in \hat{\mathcal{T}}_\tau \hskip 0.05 in \hat {\mathcal U}_{\hat H_{\trm{KS}}}(\tau,\tau_0).
\end{align}
where  $\hat{H}_{\trm{KS}}$  acts  on  the  entire
Kohn-Sham  many-body  wavefunction  $\ket{\Phi}$   through  its
constituent   single-particle  states   $\ket{\phi_j}$, see
Eq.~(\ref{eq:TDKS}).   In  general  $\lambda_{\trm{KS}}\qty(\tau)$
differs from  the constant $\lambda(\tau)$ of the interacting
system, and in addition to normalizing the many-body state, it can 
account for orthonormalization of the constituent single-particle states.

Orthonormalization
of the occupied single-particle states is an invertible transformation
as it preserves the subspace spanned by these linearly-independent states.
Representing the orthonormalization by matrix $\vb{S}$
and given a single-particle Slater
determinant wavefunction $\Phi\qty(\vb{r}_1,\vb{r}_2,\ldots,\vb{r}_N)$, it is straightforward to show that orthonormalization results in
 $\ti{\Phi}\qty({\vb{r}}_1,{\vb{r}}_2,\ldots,{\vb{r}}_N) =\Phi\qty({\vb{r}}_1,{\vb{r}}_2,\ldots,{\vb{r}}_N) \det\vb{S} $.
  Thus, the orthonormalization step  merely amounts to  changing the phase and rescaling the many-body wavefunction.

  At the starting time $\tau_0$,  we assume the Kohn-Sham wavefunction
  is properly  normalized. Following the application  of the imaginary-time
  propagator up  to a  particular  time $\tau$,  we represent  a
  particular orthonormalization  of the  single-particle states  by an
  invertible    transformation    ${\vb{S}}(\tau)$.   In    order    for
  $\lambda_{\trm{KS}}(\tau)$  to   act  like   the  orthonormalization
  procedure, we require that
\begin{align}
 \hat {\mathcal U}_{\lambda_{\trm{KS}}}(\tau,\tau_0)
  \hskip 0.05 in &= \det \vb{S}(\tau).
  \label{eq:lambdaconstraint}
\end{align}
Note  that $\abs{\det  S(\tau)}$ will  be continuous  since it  is the
reciprocal   of    the   norm    of   the    unnormalized   propagated
wavefunction. The  phase of $\det  S(\tau)$ is not important
since it changes the phase of the wavefunction, which
will  not  affect  the density.  We can therefore use  any
orthonormalization procedure  at each  time-step without  concern about
the    continuity    of    the    phase,    and    a    purely    real
$\lambda_{\trm{KS}}(\tau)$  satisfying $\hat {\mathcal U}_{\lambda_{\trm{KS}}}(\tau,\tau_0)
   = \abs{\det \vb{S}(\tau)}$ for
all $\tau > \tau_0$ is guaranteed to exist. 

The above definitions for norm-conserving $\lambda(\tau)$ and $\lambda_{\trm{KS}}(\tau)$ conclude the proof of the imaginary time extension to the van Leeuwen theorem presented in Section~\ref{ssec:vlt}.

\subsection{Monotonically Decreasing Energy} \label{ssec:monotonedecrease}

In  the  Kohn-Sham   system  the
Hamiltonian  depends on  the density,  and thus  will in  general have
eigenenergies and eigenvalues  that depend on time. In particular,
for  the   density  at   time  $\tau_\ell$,  $n({\vb{r}}, \tau_\ell)$,   we  are
considering     a    quantum     system    with     the    Hamiltonian
$\hat{H}_{\trm{KS}}[n({\vb{r}},\tau_\ell)]$.  By   propagating  the   state  of
interest in  imaginary time  using this instantaneous  Hamiltonian, we
are amplifying  the low-energy eigenstates of  the current Hamiltonian
$\hat{H}_{\trm{KS}}[n(\tau_\ell)]$,  which  in general  are  different
than   the   low-energy   eigenstates    of   the   new   Hamiltonian,
$\hat{H}_{\trm{KS}}[n(\tau_{\ell+1})]$, and the resultant state could have a higher  energy than the previous state. A  good example of this
is  the commonly-observed  divergence of  SCF loops  without a  mixing
scheme: the $N$-lowest eigenstates of  the  Hamiltonian  $\hat{H}_{\trm{KS}}[n_i]$  are directly used  to compute the  next density $n_{i+1}$. 
This  also   reveals   an interesting  limiting  case  of  it-TDDFT. 
If a KS state is propagated to infinite imaginary time before the density
used in the instantaneous Hamiltonian $\hat{H}_{\trm{KS}}[n(\tau_\ell)]$
is updated, the propagated state will become the ground state of the
present Hamiltonian, which is equivalent to populating the $N$-lowest
eigenstates of $\hat{H}_{\trm{KS}}[n(\tau_\ell)]$.
In this way basic SCF can be thought of as it-TDDFT 
  with  infinitely large  time-steps when using explicit
propagation.  Indeed, if  the
time-step  in it-TDDFT  is taken  to  be too  large, the  total energy  will
diverge, just like in SCF performed without a mixing scheme.

With  a  reasonable time-step  (usually  around  $1 \hbar/\trm{Ry}$  or
smaller), it-TDDFT monotonically  decreases the total  energy of the system.  The van
Leeuwen  theorem, which connects the KS system to the interacting system, provides  the theoretical  backbone for  this
result. While propagation  of the  Kohn-Sham system  is
complicated  by the  dependence on  density, in  the true  interacting
system the  evolution in imaginary time  has the simple form  given in
Eq.~(\ref{eq:interactingevolutionnormalized}).

\subsection{it-TDDFT as an Alternative Theoretical Foundation for Stationary States in DFT}

The first step in  the majority of DFT calculations  is to find a
density corresponding to  a stationary state. A stationary state
is an  eigenstate   of  the   Hamiltonian,  or
equivalently, a state that only  changes by a phase when evolved in
real  time or  by a  multiplicative factor  when evolved  in imaginary
time. Only the first definition is used in KS systems, as it is implicitly assumed by SCF schemes. In systems  that are  difficult to
converge  with SCF,  owing to  their size  or metallic  character, the
second definition becomes more useful, and it can be applied through the it-TDDFT method.

The  KS equations are set as an eigenvalue
problem, and thus use the first definition. Once a density $n({\vb{r}})$ is found such that a choice of
$N$ of the  single-particle eigenfunctions $\phi_j({\vb{r}})$ reproduces
the  same density  through Eq.~(\ref{eq:TDKS-density}), a  stationary
state has been determined. SCF is used to find ground states, where the $N$ lowest-energy eigenstates are
chosen,   and  $\Delta$SCF\cite{deltaSCF_ass},   where  a different  selection   of  $N$
single-particle  eigenstates   is  chosen, is used to  find  excited
states. For small systems, insulating  systems, and systems
with low  degeneracy of single-particle  states, after a few  steps of SCF, the  eigenstates rarely change  order when
sorted  by energy  from one  step  to the  next. This  means that occupied
single-particle states have similar character to
those  from the  step  before, so  the density does not change drastically. In these cases SCF converges well so using the eigenstate definition of stationary states is sensible. However, in  large systems and
in metallic  systems, or if an  excited state is desired, the above conditions might not hold, leading to charge sloshing. In  principle, the KS equations can still be used to verify a stationary state if the density is perfectly converged. In practice, this definition is inadequate in these
difficult systems  since a suitable approximate density 
could appear to  be far from convergence if  the wrong KS eigenstates
are occupied, due to the next SCF step returning a very different
density from the one given. In  addition, this makes it challenging to
determine the quality of a non-converging  density. For example,
in  Section~\ref{section:calculations}  we examine  the performance  of SCF  on a
ruthenium nanocluster, where we show that some
non-converged  densities give  a  reasonable energy  estimate for  the
ground state, while others are incorrect.

To address  convergence issues,  DFT calculations of  metallic systems
and  systems  with high  single-particle  energy  degeneracy are  often
performed with  electronic smearing\cite{Rabuck}, where states near  the Fermi
level are given fractional  occupations to simulate nonzero electronic
temperature. This  mitigates  the problem by ensuring that
states   near   each  other   in   energy   have  similar   fractional
occupation. Smearing adds an entropic contribution to the energy, so a
balance between obtaining  an accurate energy and  ease of convergence
has  to   be  struck.  Electronic  smearing   is  a
computational tool and  not intended to be  an accurate representation
of  the  effects of  temperature, so it should  be
incrementally  reduced   until  the  solution  with   no  smearing  is
achieved\cite{Michelini}. In fact, cases
have been found where even small amounts of electronic smearing 
produce significantly different results from the same calculation
performed with integer occupations, such as a HOMO-LUMO
gap energy that differ by one order of magnitude\cite{Basiuk}.
As we show in the ruthenium  nanocluster system in Section~\ref{section:calculations},  achieving
convergence while applying electronic  smearing can still require
finesse and guesswork.

In systems where  SCF convergence is hard to attain,  instead of using
the KS equations to define a stationary state, we can use a state's invariance under 
imaginary-time  evolution. If a KS wavefunction  stays constant  when
propagated in imaginary time, then its single-particle states span the
same subspace  as a  set of eigenstates which solve the KS equations. The converse  is true  as well,
namely  that  a set  of  $N$ KS eigenstates satisfying the KS equations self-consistently will be invariant under imaginary-time evolution, ignoring the possibly
changing  norm  which can  be  corrected (as discussed in Sec. \ref{ssec:orthon}). Thus, finding a  KS many-body state $\ket{\Phi(\tau_0)}$ such that $\ket{\Phi(\tau)} = \hat{\mathcal{U}}(\tau,\tau_0) \ket{\Phi\qty(\tau_0)} =
\ket{\Phi(\tau_0)}$,  where  the  Hamiltonian in the propagator   $\hat{\mathcal{U}}$   contains orthonormality-preserving
$\lambda(\tau)$, is equivalent to finding a  set of $N$
single-particle  states that  satisfy  the KS equations.

This definition has  a few
advantages. In systems  where the single-particle states are
close in energy, occupation ambiguities and charge-sloshing issues are eliminated because it-TDDFT follows the  occupied  orbitals throughout  their
evolution. Additionally, it-TDDFT handles systems with degenerate states well since an initial state will converge to one of the states within the degenerate stationary-state subspace without being affected by the unoccupied states of identical energy.

\subsection{Practical Advantages of it-TDDFT}
\label{ssec:practicaladvantages}

One convenience afforded by it-TDDFT is that  a user only  needs to
choose  a single  parameter, the time-step, when  attempting to  converge  a  system. Compare this to the various parameters usually required for
SCF with a mixing scheme: the number of past states to mix, the mixing weight, and the amount of electronic smearing, to name a few. When encountering a set of nonconvergent parameters, it is often unclear which direction to change each parameter for a better chance at convergence. In addition, there are systems where different 
stationary states  can be obtained for slight variations in the mixing parameters, as shown
in  the case  of a  Cu$_{13}$ cluster in Section~\ref{section:calculations}. In
contrast, convergence in  the it-TDDFT method is not very  sensitive to the
choice of time-step, and any choice smaller than a convergent time-step will
lead to the  same density trajectory in imaginary time  given the same
starting  state.  This  property  allows us  to eliminate this
parameter choice if desired  through algorithms that automatically
adjust the time-step on the fly. We found that the simple procedure of increasing the time-step while total energy decreases, and decreasing the time-step when it does not, can perform nearly as well as using a static convergent time-step that is as large as possible.

Another practical advantage of using imaginary-time evolution is that not-yet-converged states still have physical meaning. The single-particle states and the electronic density used in the KS Hamiltonian are self-consistent at all times, and in principle this density trajectory is equal to the imaginary-time evolving density of the interacting system by the van Leeuwen theorem. Through this connection, the partially-converged KS state corresponds to a superposition of a dominant ground state component and a few low-amplitude excited states. As such, even before the it-TDDFT ground state calculation has converged according to user-specified energy or density tolerance criteria, approximate ground state observables can be computed. This property allows for preliminary calculations of band structure, energies, optical properties, or atomic forces while the ground state calculation continues to be refined. In contrast, there are no guarantees of validity for observables calculated from intermediate states produced in a SCF loop since they are not self-consistent and can be far from the correct KS ground state.

When an SCF loop ceases to make progress, not much is gained aside from the knowledge that the particular set of mixing parameters did not lead to convergence. It could take a subtantial amount of tweaking of these parameters before chancing upon a set that works, consuming time and computational resources. This highlights another strength of it-TDDFT: the calculation time used will always improve the quality of the state at hand. It is also straightforward to continue a calculation from the last saved state, enabling incremental improvement of an approximate stationary state over multiple runs.

In our  discussion we have assumed  that we are performing  DFT with a
Kohn-Sham system, which uses a single Slater determinant. It is
possible  to  apply  it-TDDFT  for   finding  stationary  states  in  other
approaches which  use linear  combinations of Slater  determinants, or
ensemble DFT, since a DFT model that  reproduces  the  same  density
trajectory  as   the  true  interacting  system  will  evolve an arbitrary
starting state into a stationary state when propagated in 
imaginary time.

\section{Example Calculations}
\label{section:calculations}

In order  to compare two different  densities, it is useful  to have a
measure of  distance. We will
use half the $L^1$  distance for  its intuitive  physical meaning:
\begin{align}
  D[n,n_0]   \equiv    \frac{1}{2}d_1(n,n_0)   =    \frac{1}{2}   \int
  \abs{n({\vb{r}}) - n_0({\vb{r}})} \dd[3]{\vb{r}},
  \label{eq:ourdistance}
\end{align}
which can be interpreted as the number of electrons in the wrong place
relative to the reference density $n_0$. This can be seen by using the
fact that both densities integrate to the same value, the total number
of electrons. The integral  of the
absolute value  of the density difference  over all space adds  up the
excess  density  and  the  negative   of  the  lacking  density,  both
contributing equally, so the $1/2$ factor is needed to obtain the number
of electrons out of place.

\begin{figure}
  \centering
  \includegraphics[width=\figwidth]{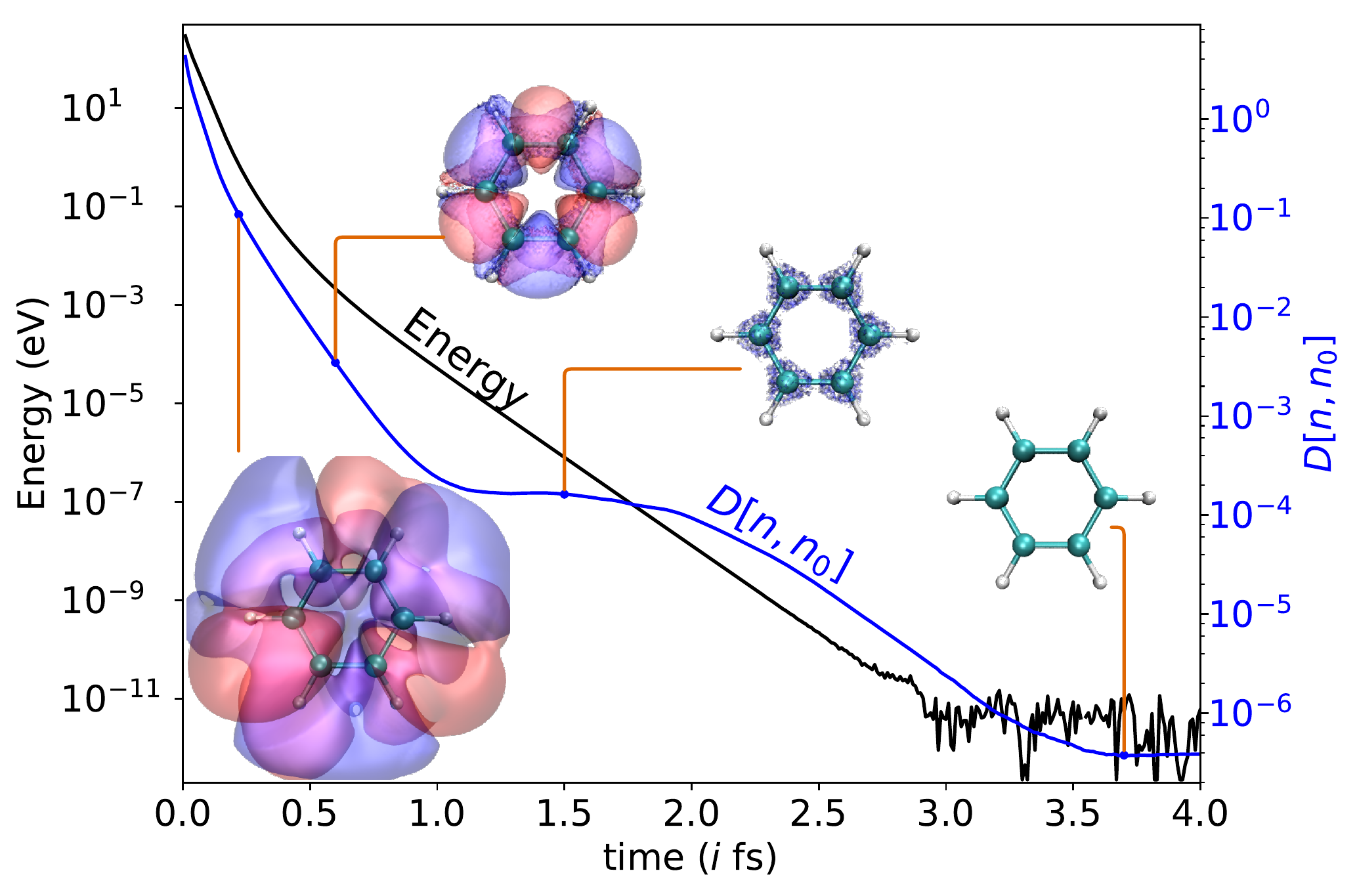}
  \caption{Determining the ground state energy $E$ and density distance $D[n,n_0]$ of benzene using it-TDDFT, relative to SCF results. The time
    step  used  was  $10.0\,\trm{as}$. Positive and  negative isosurfaces of the  density
    difference $n({\vb{r}})  - n_0({\vb{r}})$ at fixed values are shown  at various points in the propagation.}
  \label{fig:benzene}
\end{figure}

As a demonstration of using it-TDDFT to determine a ground state, we apply  the method to a benzene molecule and show that  it produces  the  same  density and  energy  as  a standard  SCF calculation. We  initialize the single-particle   electronic  states   by  drawing   basis
coefficients from a uniform distribution and orthonormalizing the  single-particle
wavefunctions. Propagating  this initial  state in imaginary  time, we
 obtain the  same  ground  state as that determined  by an  SCF
approach  with Pulay  mixing.  The Kohn-Sham total energy $E$ and the  density
distance $D[n,n_0]$ of  the  propagated  state  are plotted  as  a  function  of
imaginary time in Fig.~\ref{fig:benzene}, both relative to
the SCF-determined ground state. These quantities tend to
zero, showing that it-TDDFT indeed produces a Kohn-Sham  state that has the same energy and
density  as   the  ground  state   determined  with  an SCF
approach. As an additional check, running SCF at the end
of the  imaginary-time propagation produces SCF convergence  after the
first step.

\begin{figure}
  \centering \includegraphics[width=\figwidth]{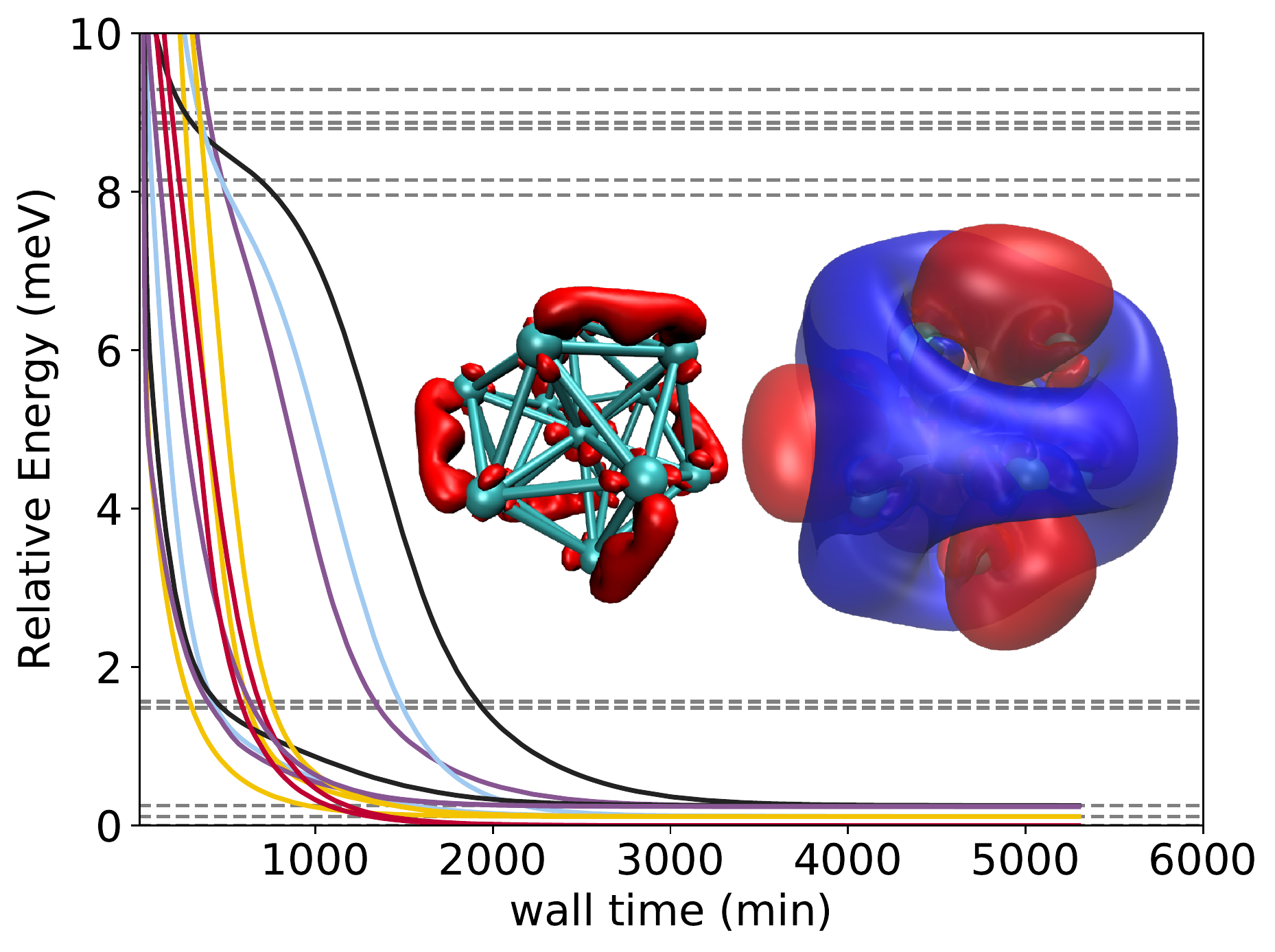}
  \caption{Electronic energy of Cu$_{13}$ with fixed spin polarization
    $+1/2$, relative  to the  lowest energy  obtained with  this fixed
    polarization. Each curve is  an energy trajectory produced by
    propagating  a random  initial  state in  imaginary time,  plotted
    versus the  wall time and colored according to the final state obtained. The right inset plot is the spin magnetization density of one such state with spin up and spin down designated by blue and red respectively. The left inset plot is a spin down isosurface to illustrate the five-fold symmetry of the lowest energy states. The horizontal dashed  lines show  the  relative energies  of
    converged  states  obtained  using   SCF  and  Pulay  mixing  with
    different parameters as detailed in Fig.~\ref{fig:barcu13} and Table~\ref{tab:cu13}.}
  \label{fig:cu13}
\end{figure}

\begin{figure*}
  \centering
  \includegraphics[width=\figwtwice]{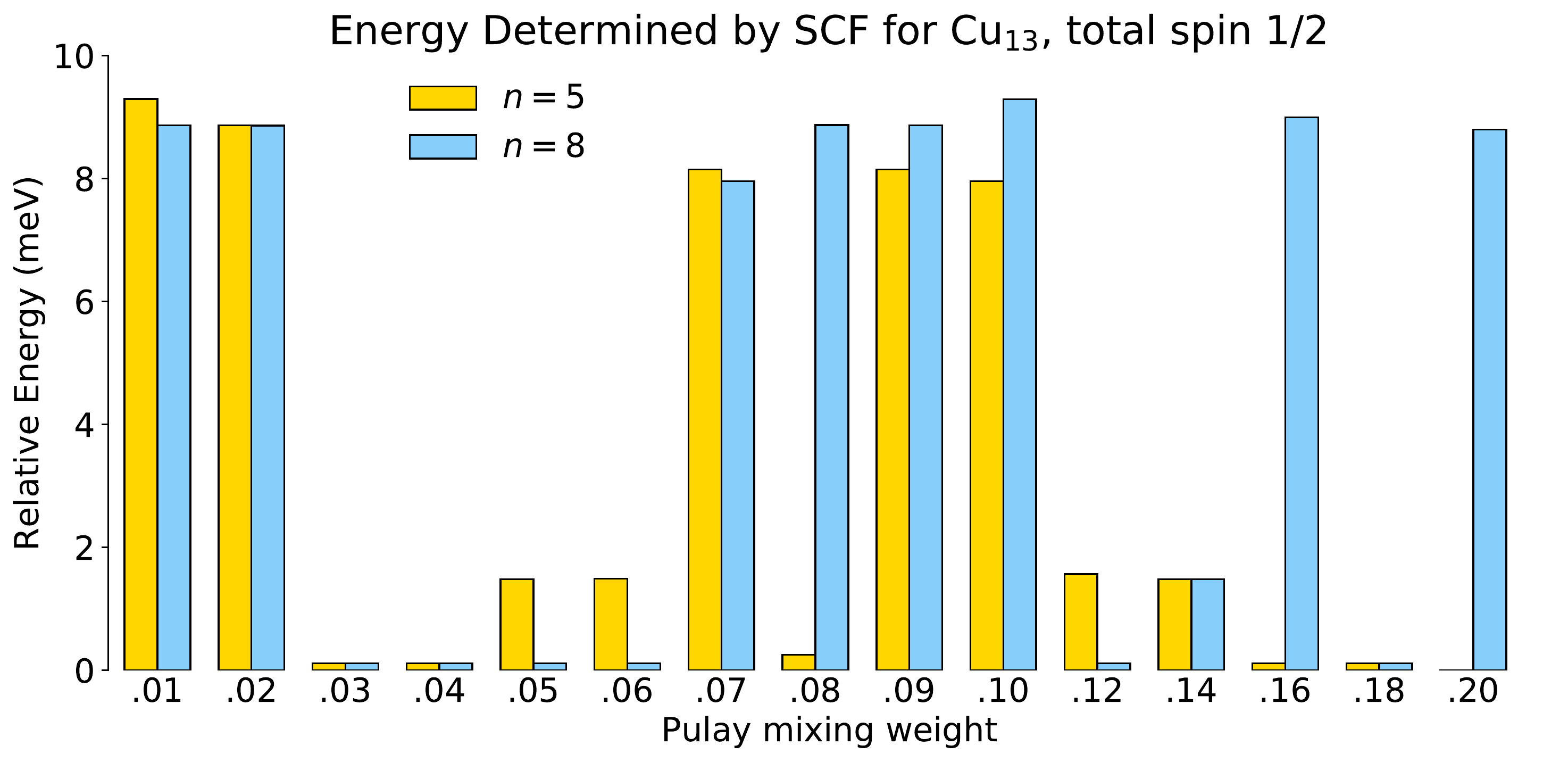}
  \caption{Relative total energy of Cu$_{13}$ cluster with fixed total spin $1/2$, obtained by SCF with
    Pulay mixing involving $n=5$ or $8$ previous densities and different mixing weights. The reference value of the energy is that obtained with imaginary-time propagation. The energies here appear in Fig.~\ref{fig:cu13} as horizontal dashed lines.}
  \label{fig:barcu13}
\end{figure*}

As our next example we consider the Cu$_{13}$ nanocluster.
Hoyt \textit{et al.}\cite{hoyt} simulated this system in its ground state magnetization of $m=5 \mu_B$ and in an excited state with magnetization
$m=3 \mu_B$, commenting that the $m= 1 \mu_B$ excited state
was tricky to converge, making it a good candidate for our it-TDDFT method. In Fig.~\ref{fig:cu13} and \ref{fig:barcu13}, we present the main results of our computations for the self-consistent KS states with $m = 1 \mu_B$ magnetization, which has total spin $1/2$. Additional information can be found in Table~\ref{tab:cu13}.
SCF has trouble finding  the minimum energy states in this fixed-spin system, due to  the fact that there
are  five degenerate  states\cite{hoyt}. Fig.~\ref{fig:cu13} shows  the energy
trajectories  in  imaginary  time  of  12  different  random  starting
configurations, and  each of  these converges  to one of five
lowest-energy states. To help identify the equality of final states, for both the it-TDDFT and SCF runs, we also computed the density distances between each combination of  obtained states. 
States with energies within $10^{-2}\,\textrm{meV}$ and a density distance of less than $(1/100)e$ of each other were considered equal. The ground state of the system, with magnetization $m = 5 \mu_B$, contains five degenerate valence electrons which have unpaired spins\cite{hoyt}. To obtain a magnetization of $m = 1 \mu_B$, four of the electrons need to pair spins, leaving five possibilities of which electron remains unpaired. Fig.~\ref{fig:cu13} shows the magnetization density, defined as the difference between spin up and spin down electron density, of one of these five lowest-energy states. There are five equivalent ways to place such a magnetization density on the icosahedral shape of the copper cluster, explaining the degeneracy. The visible differences of up to $0.1\,\textrm{meV}$ in the energies of these degenerate states are due to the discretization effects of the real space grid breaking the icosahedral symmetry. 

Our approach is better at finding the lowest-energy
states compared to SCF, which for different mixing
parameters often  converges to other excited  states of spin $+1/2$ (the  energies of
which are shown as dashed lines in the figure). In Fig.~\ref{fig:barcu13}, we show the electronic energies of the states obtained using SCF with Pulay mixing, for various mixing parameter choices. Even small changes in the mixing parameters can result in a different final state. This happens in metallic systems where the gap between occupied and unoccupied states is small, causing SCF to find a low-lying excited state.

For our final example of applying our method we consider the Ru$_{55}$ nanocluster.
Montemore \textit{et al.}\cite{montemore} studied catalysis on the surface of this structure, and found that the spin-unpolarized ground state calculation was difficult to converge with SCF. 

\begin{figure}
  \centering
  \includegraphics[width=\figwidth]{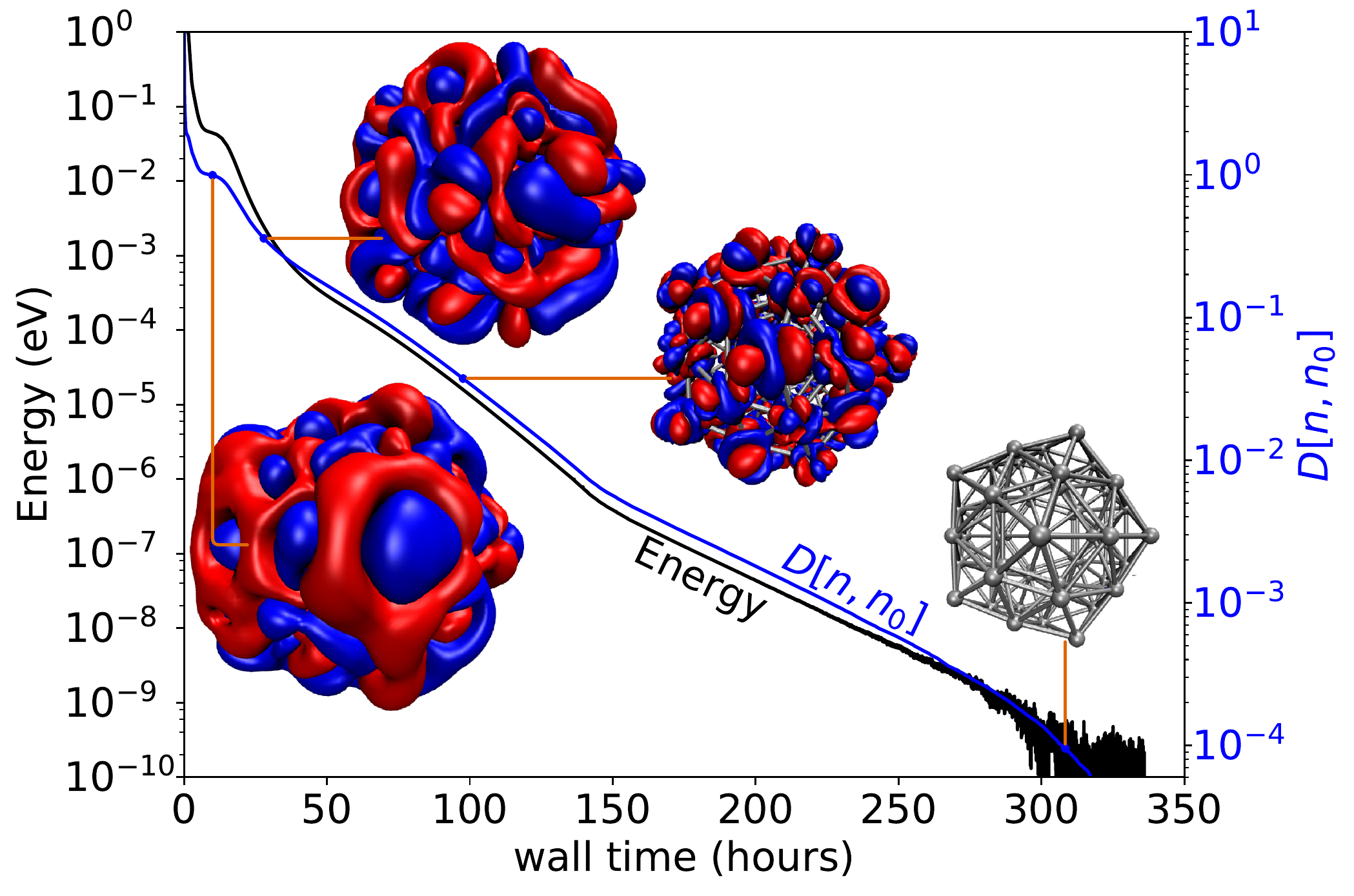}
  \caption{Electronic energy and density distance $D[n,n_0]$ trajectory of a spin unpolarized Ru$_{55}$ cluster measured relative to the state it converges to, as obtained by it-TDDFT. Positive and negative isosurfaces of $n({\vb{r}})  - n_0({\vb{r}})$ are shown at various points in the propagation.}
  \label{fig:ru55}
\end{figure}

In Table~\ref{tab:ru55}, we show the results of using SCF with Pulay mixing to find the ground state of the spin-unpolarized Ru$_{55}$ cluster. The number of past densities to mix was kept at $n = 5$ for all trials and mixing weights ranging from $0.02$ to $0.20$ were tested. We used Fermi electronic smearing for half the trials with $T = 300\,\textrm{K}$. For each run, we list the energy of the final step relative to the energy calculated with it-TDDFT, the density difference $\Delta \rho_{\textrm{max}}$, and whether the run converged or not. The density difference $\Delta \rho_{\textrm{max}}$ refers to the maximum elementwise difference in the density matrix between the final and penultimate step and is typically used to determine convergence. We used the criterion $\Delta \rho_{\textrm{max}} < 10^{-6}$. Only a few mixing weights result in convergence, namely the smallest ones with $T = 300\,\textrm{K}$ of smearing. In these runs, the entropic energy contribution is $78\,\textrm{meV}$ relative to the ground state energy. None of the runs without electronic smearing converge, despite the fact that some parameter configurations obtain energies similar to the ground state energy. The states resulting from unconverged runs generally should not be trusted as they may not be acceptable approximations to actual solutions, which have to satisfy the KS equations self-consistently. For example, in Table~\ref{tab:ru55}, examining the row with mixing weight $0.10$ and comparing the $T = 0\,\textrm{K}$ and $T = 300\,\textrm{K}$ cases, we find that even though $\Delta \rho_\textrm{max} = 0.589$ in the latter run is smaller than $\Delta \rho_\textrm{max} = 0.702$ in the former, the energy of the state obtained with $T = 300\,\textrm{K}$ is more than $6\,\textrm{eV}$ off from the correct ground state energy while the $T = 0\,\textrm{K}$ run is only about $0.006\,\textrm{eV}$ off. Applying our it-TDDFT method to the Ru$_{55}$ cluster produces the ground state without issue, as illustrated in Fig.~\ref{fig:ru55}. The observed monotonically decreasing energy and density distance $D[n,n_0]$ show consistent progress, as we expect from the theory.

\begin{table*}
  \centering
  {\textbf{SCF with Pulay Mixing for Ru$_{55}$, Spin Unpolarized} } \\
  \begin{tabular}{c|rrr|rrr}
    \toprule[1.5pt]
    & \multicolumn{3}{c|}{$T = 0\,\textrm{K}$}  & \multicolumn{3}{c}{$T = 300\,\textrm{K}$} \\
     Weight & Energy (meV) & $\Delta \rho_{\trm{max}} \,$ & Converged  & Energy(meV) & $\Delta \rho_{\trm{max}} \;$ & Converged \\
    \midrule[1pt]
    $0.02$ & $-0.012 \phantom{000}$ & $0.086$ &  No  & $78.388\phantom{0}$ & $9.9 \times 10^{-7}$ & Yes \\
    $0.04$ & $-0.004\phantom{000}$ & $0.052$ & No  & $78.380\phantom{0}$ & $9.8 \times 10^{-7}$ & Yes \\
    $0.06$ & $0.069\phantom{000} $ & $0.227$ & No  & $78.448\phantom{0}$ &$5.3 \times 10^{-7}$ & Yes \\
    $0.08$ & $0.395\phantom{000} $ & $0.185$ & No  & $78.408\phantom{0}$ & $9.0 \times 10^{-7}$ & Yes \\
    $0.10$ & $5.538\phantom{000}$ & $0.702$ & No & $6.38 \times 10^{3}\phantom{0}$ & $0.589$ & No \\
    $0.12$ & $73.276\phantom{000}$ & $0.730$ & No  & $7.15 \times 10^{4}\phantom{0}$ & $0.699$ & No \\
    $0.14$ & $148.995\phantom{000}$ & $1.388$ & No  & $1.46 \times 10^{5}\phantom{0}$ & $1.391$ & No \\
    $0.16$ & $6.385\phantom{000}$ & $0.589$ & No &  $2.35 \times 10^{5}\phantom{0}$ & $1.409$ & No \\
    $0.18$ & $295.051\phantom{000}$ & $1.441$ & No  & $2.95 \times 10^{5}\phantom{0}$ & $1.446$ & No \\
    $0.20$ & $353.573\phantom{000}$ & $1.444$ & No  & $3.54 \times 10^{5}\phantom{0}$ & $1.455$ & No \\
    \bottomrule[1.5pt]
  \end{tabular}

  \caption{Ground state electronic configurations of a Ru$_{55}$ nanocluster using SCF with Pulay mixing, with a $n=5$ density history length and mixing weights ranging from $0.02$ to $0.2$. $\Delta \rho_{\textrm{max}}$ is the maximum elementwise difference in the density matrix between the final and penultimate step, with convergence criterion $ \Delta \rho_{\textrm{max}} < 10^{-6}$.}
      \label{tab:ru55}
\end{table*}

\section{Conclusion}
\label{section:conclusion}

The first step of any Kohn-Sham DFT calculation is the determination of a self-consistent solution to the KS equations, resulting in a density corresponding to a stationary state of the many-body interacting system. While the standard method of using the iterative SCF procedure generally produces a solution efficiently, there are important classes of systems that pose problems for this approach due to their small band gaps or degenerate single-particle energies. We have proposed the it-TDDFT method as an alternative means for solving the KS equations in these difficult systems, and shown how it avoids the issues which affect SCF. 

We established that the van Leeuwen theorem, a key theoretical foundation for TDDFT methods, can be extended to imaginary time, thereby ensuring convergence to a stationary state independent of the exchange-correlation potential and level of theory used in the model system. In addition, we discussed how it-TDDFT could be used in an alternative but equivalent definition of stationary states in DFT, better suited for metallic systems and systems with degenerate or nearly-degenerate states and based on the time-dependent Kohn-Sham equations. The it-TDDFT method also exhibits a number of practical advantages, such as justifying approximations to observables of interest before the ground state calculation is fully converged, requiring few input parameters, and allowing easy refinements of the results of previous runs by continuing from a saved state. 

In the copper and ruthenium nanoclusters considered here, we demonstrated how SCF can struggle to find the electronic ground state, either converging to low-lying excited states or getting stuck in charge-sloshing cycles. These systems were readily converged by it-TDDFT, showcasing its robustness through smooth trajectories with monotonically decreasing energy. For these systems we either ran the calculation as spin-unpolarized, or with a fixed total spin. This is not an inherent limitation of the method, as one could simply run the calculation with all possible spin polarizations and select the state with the lowest energy. The method can be adapted to non-collinear spin systems, since the operating principle depends only on the Hamiltonian being able to differentiate states by energy. Furthermore, while we used finite systems for our example calculations, our method can be extended to find ground states of periodic systems by simultaneously propagating Kohn-Sham states at multiple $k$-points.

Given an existing TDDFT code which evolves systems in real-time, it should be relatively straightforward to implement a prototype of the presented it-TDDFT approach, requiring only an imaginary time substitution in the propagation step and a method to orthonormalize the single-particle states. While more efficient implementations could be examined in the future, the low barrier to utilizing it-TDDFT could make it an attractive alternative option for those dealing with particularly vexing systems.

\FloatBarrier
\bibliography{references_nourl}

\section{Supplemental Materials}
\label{section:supplemental}

\begin{table}[h]
  \centering  {\textbf{States  Determined  by  SCF with  Pulay  Mixing  for
    Cu$_{13}$, total spin $1/2$ }} \\
  \begin{tabular}{l|rrr|rrrr}
    \toprule[1.5pt]
    & \multicolumn{3}{c|}{$n=5$} & \multicolumn{3}{c}{$n=8$} \\
    Mixing & \multicolumn{1}{c}{Energy }  &\multicolumn{1}{c}{ Time }&\multicolumn{1}{c|}{ Min. }&\multicolumn{1}{c}{ Energy }&\multicolumn{1}{c}{ Time }&\multicolumn{1}{c}{ Min.}\\
    Weight &\multicolumn{1}{c}{ (meV) }&\multicolumn{1}{c}{ (min) }&\multicolumn{1}{c|}{ State }&\multicolumn{1}{c}{ (meV) }&\multicolumn{1}{c}{ (min) }&\multicolumn{1}{c}{ State}\\
    \midrule[1pt]
$0.01$ & $9.2955 $&$2021.7$&	No  &   $8.8701$ & $787.8$&   No \\
$0.02$ & $8.8690 $&$2386.0$&	No  &   $8.8656$ & $750.6$&   No \\
$0.03$ & $0.1155 $&$18.7$&      Yes &   $0.1135$ & $26.8$&   Yes \\
$0.04$ & $0.1156 $&$51.0$&	Yes &   $0.1162$ & $14.9$&   Yes \\
$0.05$ & $1.4833 $&$30.7$&	No  &   $0.1164$ & $15.4$&   Yes \\
$0.06$ & $1.4910 $&$64.1$&	No  &   $0.1176$ & $13.9$&   Yes \\
$0.07$ & $8.1505 $&$163.4$&	No  &   $7.9599$ & $29.3$&   No \\
$0.08$ & $0.2538 $&$28.0$&      Yes &   $8.8722$ & $119.4$&   No \\
$0.09$ & $8.1506 $&$61.2$&	No  &   $8.8698$ & $207.5$&   No \\
$0.10$ & $7.9600 $&$60.3$&	No  &   $9.2955$ & $137.4$&   No \\
$0.12$ & $1.5656 $&$22.8$&      No  &   $0.1159$ & $17.9$&   Yes \\
$0.14$ & $1.4832 $&$34.3$&      No  &   $1.4844$ & $15.3$&   No \\
$0.16$ & $0.1163 $&$7.0$&	Yes &   $8.9980$ & $32.0$&   No \\
$0.18$ & $0.1170 $&$6.8$&	Yes &   $0.1162$ & $10.7$&   Yes \\
$0.20$ & $0.0003 $&$7.4$&	Yes &   $8.8015$ & $54.8$&   No \\
    \bottomrule[1.5pt]
  \end{tabular}
  \caption{Electronic states of spin $1/2$ Cu$_{13}$ obtained from SCF with no electronic smearing. For the Pulay mixing weight and number of past densities $n$ used in the mixing scheme, we list the relative energy, wall time, and whether the converged state matches one of the five lowest-energy states determined with it-TDDFT. The wall times can be compared to those shown in Fig.~\ref{fig:cu13}.}
  \label{tab:cu13}
\end{table}
\end{document}